\documentclass[journal]{IEEEtran}
\IEEEoverridecommandlockouts
\usepackage{algorithmic}
\usepackage{graphicx}
\usepackage{textcomp}
\usepackage{nopageno}
\usepackage{xcolor, soul}
\usepackage{amsmath, amsfonts, amssymb, amsthm}
\usepackage{color,soul}
\usepackage{graphicx}
\usepackage{fancyhdr}
\usepackage{array}
\usepackage{upgreek}
\usepackage{soul}
\usepackage{subcaption}
\usepackage{longtable}
\usepackage{tabularray}
\usepackage{hyperref}
\usepackage{multicol}
\usepackage{authblk}
\usepackage{cite}
\usepackage{setspace}

\sethlcolor{green}

\usepackage{svg}
\usepackage{booktabs}
\usepackage{float}
\def\BibTeX{{\rm B\kern-.05em{\sc i\kern-.025em b}\kern-.08em
    T\kern-.1667em\lower.7ex\hbox{E}\kern-.125emX}}

\title{Waveguide QED Analysis of Quantum-Coherent Links for Modular Quantum Computing
\thanks{*Equally contributing authors. Authors acknowledge funding from the EC through HORIZON-EIC-2022-PATHFINDEROPEN-01-101099697 (QUADRATURE) and HORIZON-ERC-2021-101042080 (WINC). CGA acknowledges support from the Spanish Ministry of Science, Innovation and Universities through the Beatriz Galindo program 2020 (BG20 00023) and the European ERDF under grant PID2021-123627OB-C51. EA acknowledges support from Generalitat de Catalunya, ICREA Academia Award 2024.}}

\author{
Junaid Khan$^{\ast\dagger}$, Sergio Navarro Reyes$^{\ast\dagger}$, Sahar Ben Rached$^{\ast\dagger}$, Eduard Alarcón$^{\dagger}$,\\
Peter Haring Bolívar$^{\ddagger}$, Carmen G. Almudéver$^{\S}$, and Sergi Abadal$^{\dagger}$}
\affil{$^{\dagger}$\textit{Universitat Politècnica de Catalunya, Spain} \\
       $^{\ddagger}$\textit{University of Siegen, Germany} \\
       $^{\S}$\textit{Universitat Politècnica de València, Spain} \\
       \text{junaid.khan@upc.edu}}

\date{}  

\begin{document}
\maketitle
\thispagestyle{empty}
\begin{abstract}
Waveguides potentially offer an effective medium for interconnecting quantum processors within a modular framework, facilitating the coherent quantum state transfer between the qubits across separate chips. In this work, we analyze a quantum communication scenario where two qubits are connected to a shared waveguide, whose resonance frequency may match or not match that of the qubits. Both configurations are simulated from the perspective of quantum electrodynamics (QED) to assess the system behavior and key factors that influence reliable inter-chip communication. The primary performance metrics analyzed are quantum state transfer fidelity and latency, considering the impact of key system parameters such as the qubit-waveguide detuning, coupling strength, waveguide decay rate, and qubit decay rate. We present the system design requirements that yield enhanced state transmission fidelity rates and lowered latency, and discuss the scalability of waveguide-mediated interconnects considering various configurations of the system.
\end{abstract}
\section{Introduction}
Quantum computing is a rapidly advancing technology, potentially capable of solving complex problems found to be intractable with classical computers \cite{gill2024quantum}. The computational power of quantum computers increases exponentially with the number of qubits \cite{gupta2020quantum}. However, integrating a large number of qubits within a single processor introduces significant challenges such as crosstalk and degrading effects of qubit decoherence, which deteriorate the system performance \cite{altomare2010measurement}. Mitigating these issues has driven the development of modular quantum processors \cite{bravyi2022future, rodrigo2021scaling}, where qubits are distributed across multiple processor \textit{cores} that are linked via quantum-coherent interconnects. The cores can be integrated within a single or multiple chips, like classical multicore processors \cite{li2020chiplet}. Either way, this approach allows to increase the computational power of the resulting modular quantum computer, yet only as long as the interconnects are able to achieve quantum state transfers with enough fidelity and speed \cite{escofet2023interconnect}.

\begin{figure}[H]
    \centering    \includegraphics[width=0.48\textwidth]{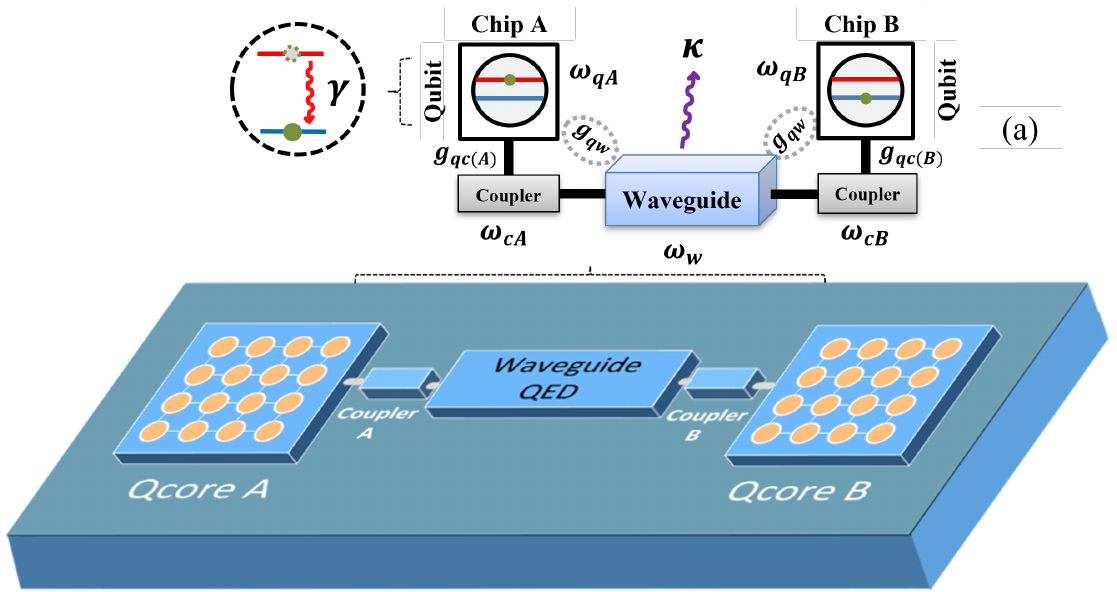} \vspace{-0.1cm} 
    \caption{Waveguide-mediated interconnect between qubits of two different chips in a modular quantum computer.}\vspace{-0.1cm}
    \label{fig:wQED-new}
\end{figure}

Waveguides could play a significant role for building inter-chip communication channels between multicore quantum processors, exploiting the Quantum Electrodynamics (QED) that occur within the waveguide (Fig.~\ref{fig:wQED-new}) \cite{sheremet2023waveguide, almanakly2022towards}. In waveguide QED systems, qubits are coupled to waveguides, which act as channels for photon-mediated quantum state transfer \cite{sanchez2020theory}. This design enhances chip-to-chip quantum communication with reduced losses and improved fidelity, enabling efficient scaling of quantum architectures to accommodate larger qubit numbers \cite{arrazola2014quantum}. These media support high-fidelity quantum operations such as qubit-to-qubit entanglement and remote gate execution, which are vital for low-latency, high-efficiency quantum state transmission between processors \cite{ferrara2021computational}.

Nevertheless, integrating an chip-to-chip quantum communication network introduces further challenges, including channel losses and qubit state transfer longer latencies, resulting in degrading the computation process. For the purpose of building robust and reliable quantum networks, various proposals for quantum interlinks are emerging, including cavity-mediated systems \cite{krasnok2024superconducting, vaidya2018tunable, srinivasa2024cavity} and electro-optic transduction \cite{delaney2022superconducting, shen2024traveling, d2023towards}. These proposals address the main obstacles facing scalability of quantum networks for modular quantum computing systems: low efficiency in state generation, transmission, and transduction \cite{yang2023survey}. The current proposals for cavity-mediated frameworks face significant challenges due to high signal losses, limiting their efficiency and scalability \cite{rached2024benchmarking}, yet research is ongoing to address these limitations by developing more efficient platforms. 

Since waveguide-mediated channels offer a promising approach for the creation of large-scale quantum computers, ongoing efforts research aim at optimizing these systems \cite{brehm2021waveguide}. However, an exploration of the potential of these channels from an engineering perspective is missing in the literature. In particular, our aim is to study the waveguide-mediated channels from an engineering perspective, seeking to relate their performance (in the form of fidelity and latency) with the
performance of the underlying components (the qubits and waveguide) as well as their interrelation.
 
To bridge this gap, this paper assesses the QED of a simple system where a waveguide mediates the transfer between qubits of two separate chips (Fig.~\ref{fig:wQED-new}). The study considers that the resonance frequency of the waveguide may match or not match with that of the qubits, to evaluate the impact of unintended detuning (due to fabrication non-idealities) or deliberate detuning (to miniaturize the waveguide) on the communication performance. This is further described in Sec.~\ref{sec:method}. Simulations assess the dynamics of each setup, characterizing the quantum state transfer fidelity and latency as critical indicators of the efficiency and scalability of the proposed systems, as presented in Sec.~\ref{sec:results}. The results are further discussed in Sec.~\ref{sec:discussion} and the paper is concluded in Sec.~\ref{sec:conclusion}.

\section{Modeling of a waveguide-mediated quantum interconnect for modular quantum processors} \label{sec:method}
For the sake of simplicity, we consider a modular quantum computer with two single-qubit chips based on electron spin qubits \cite{saraiva2022materials}, each linked to the common waveguide through tunable couplers as illustrated in Fig. \ref{fig:wQED-new}. Next, we describe the system model in more detail in Sec.~\ref{sec:dynamics} and the simulation methods and scenarios in Sec.~\ref{sec:scenarios}. The result of the study is the latency and fidelity of the quantum state transfer.

\subsection{Qubit-waveguide quantum system model}
\label{sec:dynamics}
We analyze a quantum system consisting of two single-qubit chips (A and B) interconnected to a common waveguide through a tunable coupler. The waveguide is operating at a qubit frequency of 6 GHz \cite{krantz2019quantum, veldhorst2014addressable}. In our study, we focus on the direct interaction between a qubit and the waveguide, hence considering an ideal coupler on both ends. This approach allows us to examine quantum state transfer through both a tuned and detuned waveguide, focusing on the role of the qubit and waveguide parameters in determining the efficiency of photon-mediated interactions. In this context, the system's behavior is described by the following Hamiltonian,
\begin{equation}
    \begin{aligned}
        H &= \hbar \omega_{q(A,B)} \left( \hat{\sigma}_A^{+} \hat{\sigma}_A^{-} + \hat{\sigma}_B^{+} \hat{\sigma}_B^{-} \right)
         + \hbar \omega_w \hat{a}^{\dagger} \hat{a} +\\
        &\quad + \hbar g_{qw} \left( \hat{\sigma}_A^{+} \hat{a} + \hat{\sigma}_A^{-} \hat{a}^{\dagger} + \hat{\sigma}_B^{+} \hat{a} + \hat{\sigma}_B^{-} \hat{a}^{\dagger} \right),
    \end{aligned}
    \label{Ht}
\end{equation}
where $\hbar$ is the reduced Planck's constant, $\omega_q$ is the qubit frequency and $\omega_w$ is the waveguide frequency. Additionally, the operators $\hat{\sigma}^{+}$ and $\hat{\sigma}^{-}$ represent the raising (excitation) and lowering (de-excitation) operators for a two-level quantum system (such as a qubit), while $\hat{a}^{\dagger}$ and $\hat{a}$ represent the creation and annihilation operators of the quantized electromagnetic field in the waveguide, and $g_{qw}$ is the qubit-waveguide coupling strength factor \cite{reiserer2015cavity}. 

To model decoherence and the dissipation in the quantum system, we employ the Lindblad master equation, which describes the time evolution of the system's density matrix $\rho$. This framework captures both unitary dynamics, governed by the Hamiltonian, and non-unitary effects due to environmental interactions. The equation can be expressed as 
\begin{equation}
    \begin{aligned}
     \frac{d\rho}{dt}& = -\frac{i}{\hbar} 
        [H, \rho] 
       + \gamma_q \sum_{j=A,B} \left( \hat{\sigma}_j \rho \hat{\sigma}_j^\dagger - \frac{1}{2} \{\hat{\sigma}_j^\dagger \hat{\sigma}_j, \rho\} \right) + \\
       &+ \kappa \left( \hat{a} \rho \hat{a}^\dagger - \frac{1}{2} \{\hat{a}^\dagger \hat{a}, \rho\} \right),
    \end{aligned}
\end{equation}
where $\gamma_q$ and $\kappa$ are Lindblad operators representing qubit decay rate and waveguide decay rate, respectively \cite{manzano2020short}. These terms characterize noise-induced effects on quantum state transfer, impacting fidelity and latency.

\begin{figure*}[t]
    \centering
    \begin{subfigure}[t]{0.32\textwidth}
        \centering
        \includegraphics[width=\columnwidth]{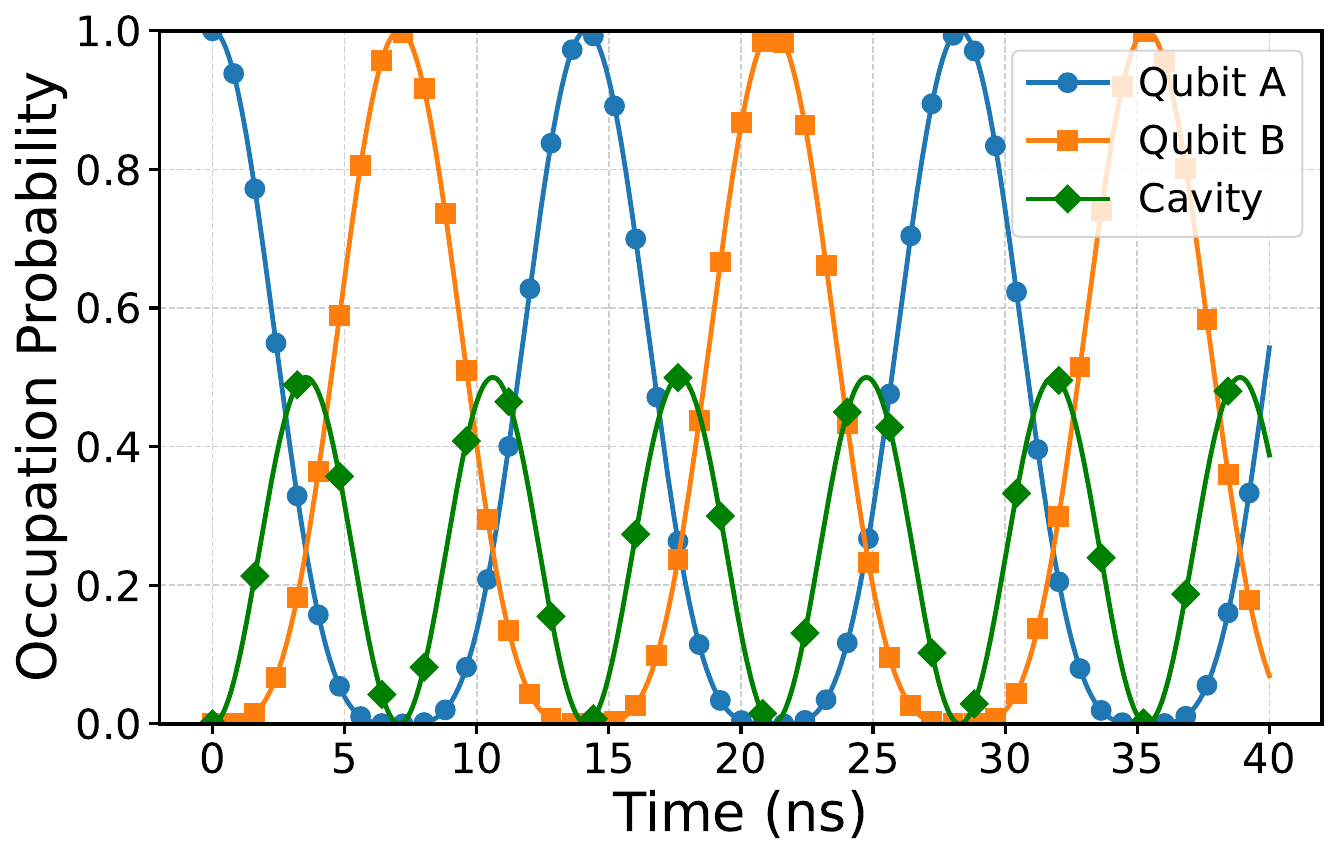} 
        \caption{\centering Lossless setup with coupling strength $g_{qw}= 0.05$ GHz.}
       \label{lossless-config} 
    \end{subfigure}
    \hfill
    \begin{subfigure}[t]{0.32\textwidth}
        \centering
        \includegraphics[width=\columnwidth]{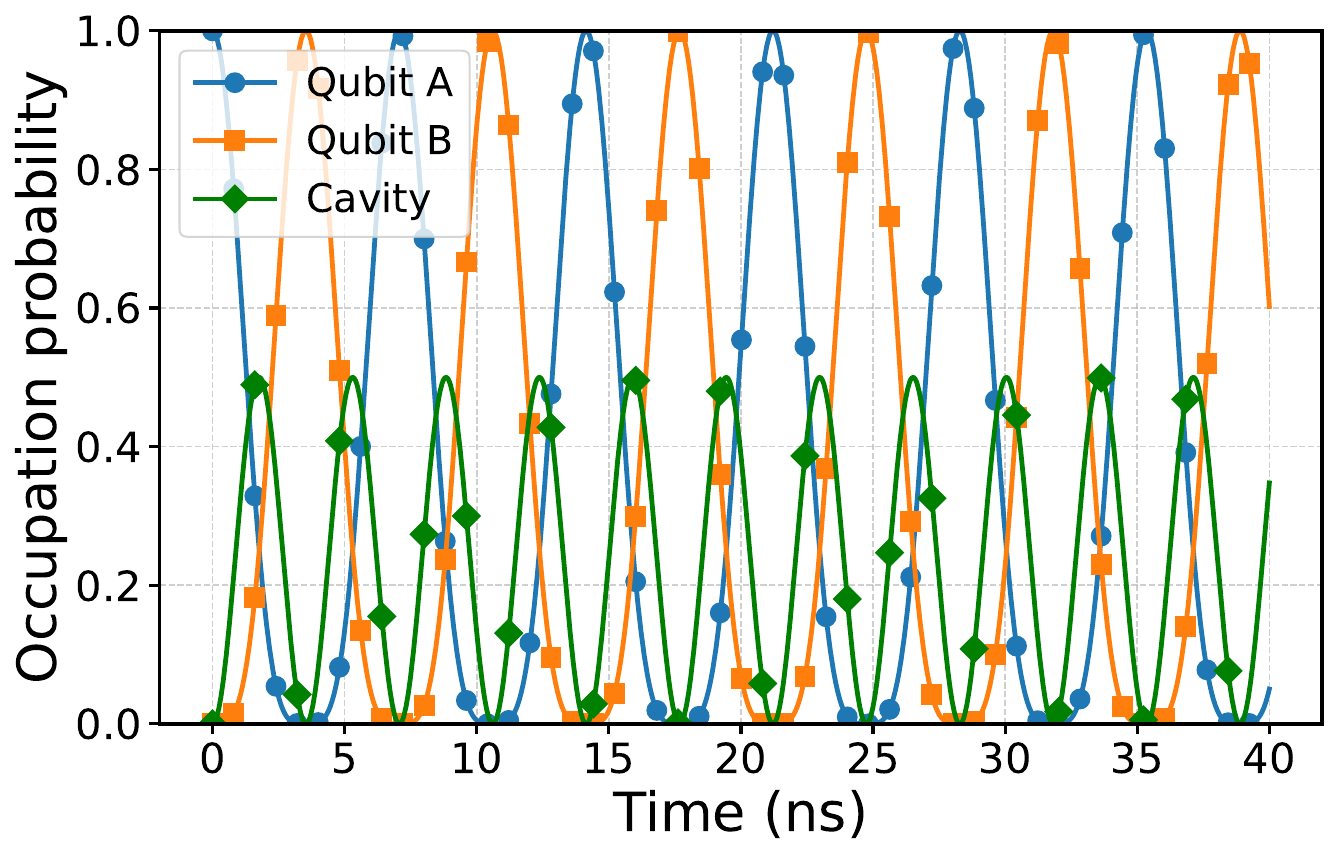} 
        \caption{\centering Lossless setup with higher coupling strength value $g_{qw}= 0.1$ GHz.}
    \label{g-increased-config}    
    \end{subfigure}
     \hfill
    \begin{subfigure}[t]{0.32\textwidth}
        \centering
        \includegraphics[width=\columnwidth]{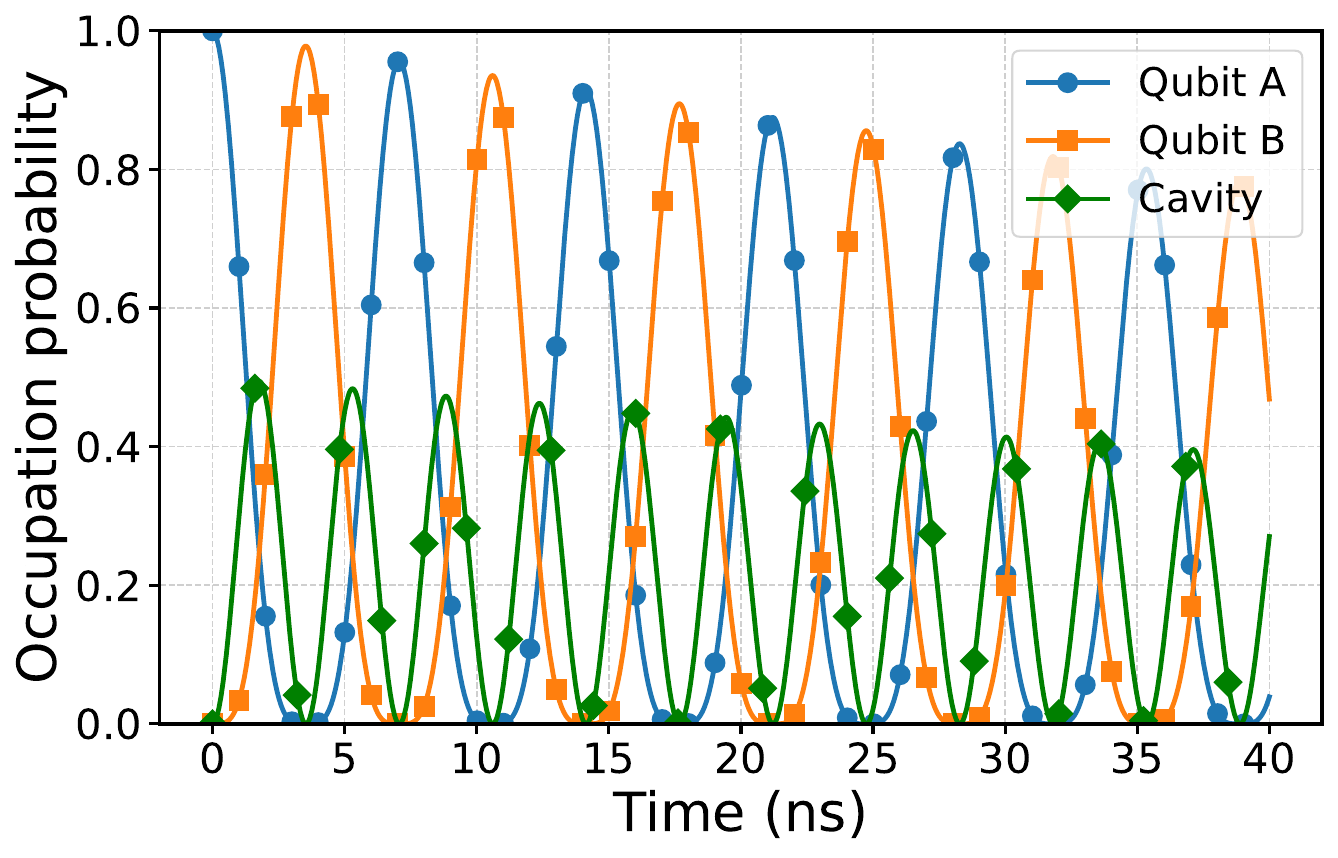} 
        \caption{\centering Lossy setup with $\kappa$, $\gamma = 1$ MHz, and high coupling strength $g_{qw}= 0.1$ GHz.}
    \label{lossy-config}    
    \end{subfigure}   
    \caption{Temporal evolution of the quantum state transfer between two qubits in the resonant case ($\omega_q = \omega_w = 6$ GHz).}
    \label{fig:Time evolution and state manipulation (Resonant prototype)}
\end{figure*}

\begin{figure*}[t]
    \centering
        \begin{subfigure}[t]{0.32\textwidth}
        \centering
        \includegraphics[width=\columnwidth]{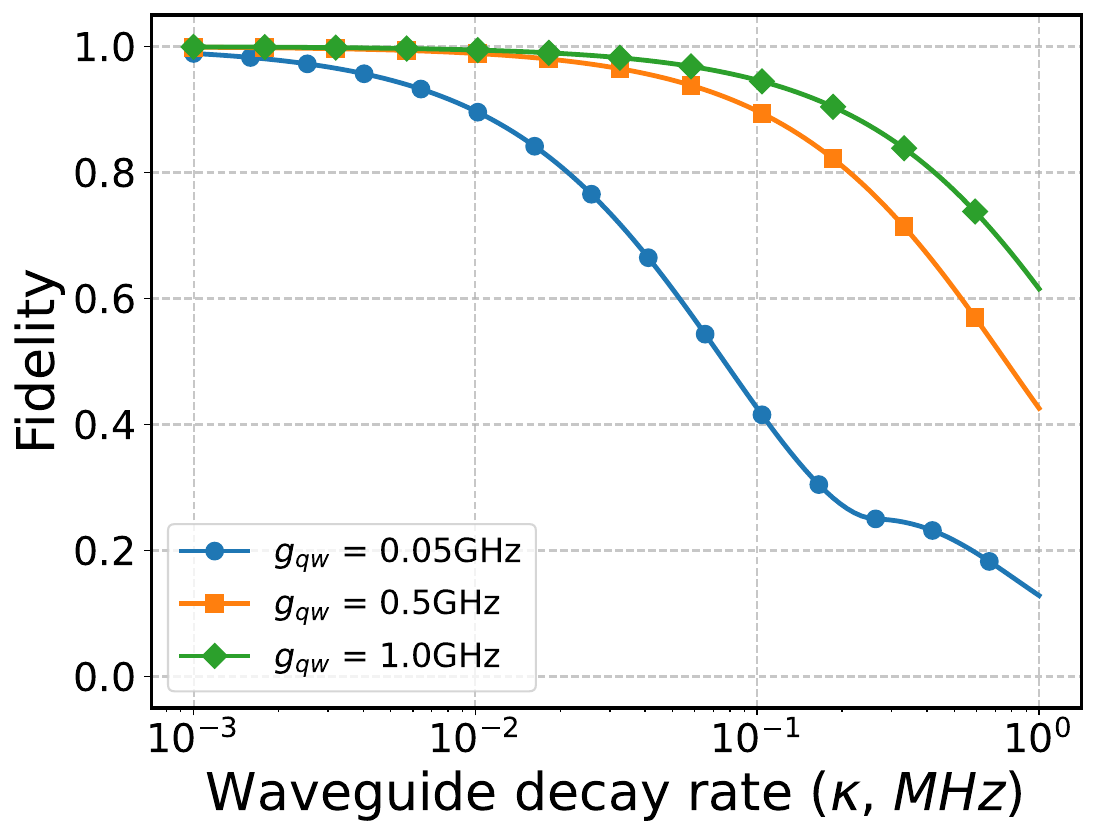}
        \caption{\centering Fidelity over the waveguide decay rate for different coupling strengths ($\gamma =$ 0).}
    \label{resonant-waveguide-decay resonant}    
    \end{subfigure}
    \begin{subfigure}[t]{0.32\textwidth}
        \centering
        \includegraphics[width=\columnwidth]{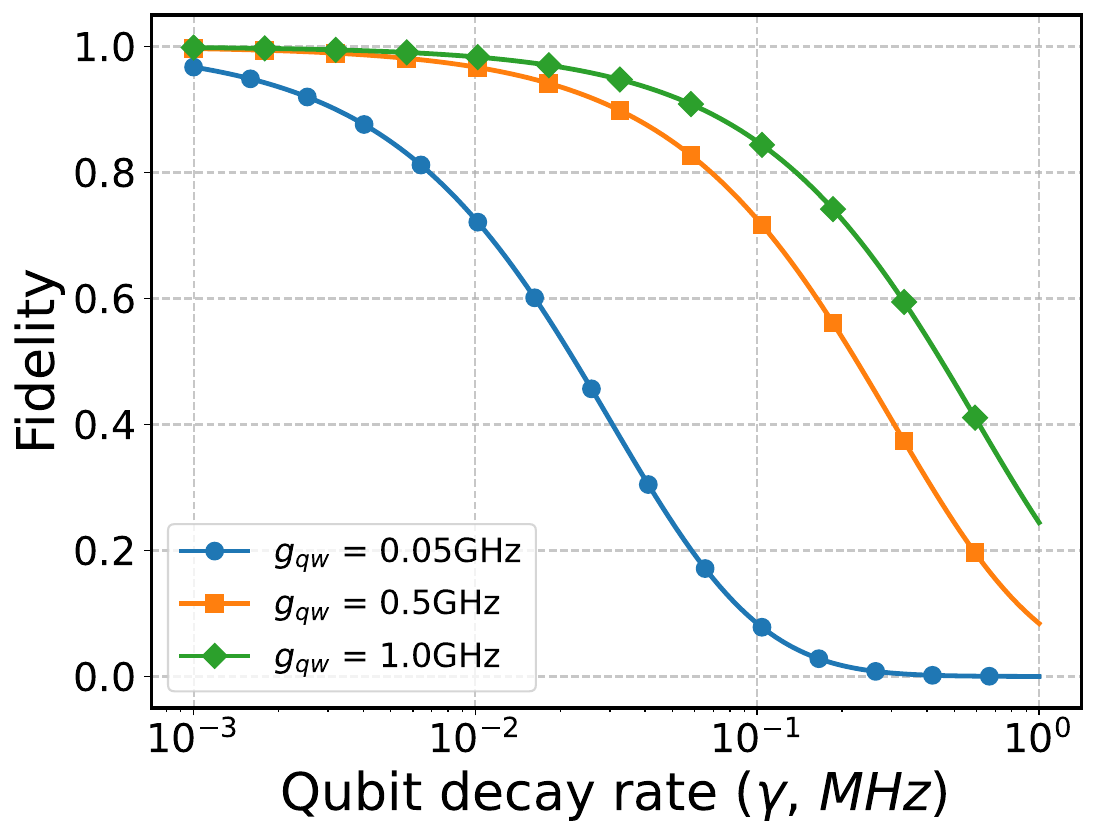}
        \caption{\centering Fidelity over the qubit decay rate for different coupling strengths ($\kappa =$ 0).}
    \label{resonant-qubit-decay resonant}    
    \end{subfigure}
    \begin{subfigure}[t]{0.32\textwidth}
        \centering
        \includegraphics[width=\columnwidth]{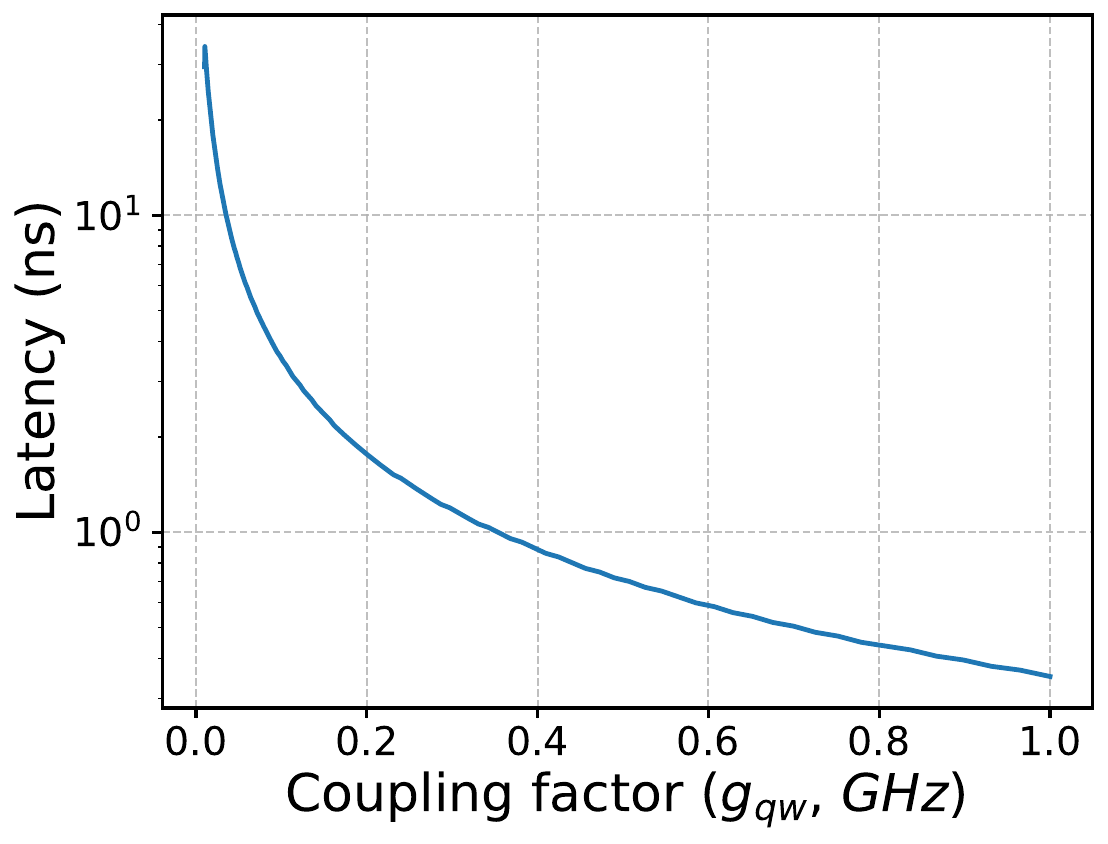}
        \caption{\centering Latency over the coupling strength ($\kappa = \gamma =$ 0).}
       \label{latency-resonant} 
    \end{subfigure}
    \caption{Assessment of the quantum state transfer performance in the resonant case ($\omega_q = \omega_w = 6$ GHz).}
    \label{res-latency-fidelity}
\end{figure*}
\subsection{Simulation methods and scenarios}
\label{sec:scenarios}
The system model outline in the previous section is simulated using QuTip \cite{johansson2012qutip}, an open-source Python package for simulating quantum systems. As outlined in Table \ref{tab:parameters}, our study considers a broad range for the key input parameters of the system, i.e. the qubit decay rate $\gamma$, the waveguide decay rage $\kappa$, and the qubit-waveguide coupling strength $g_{qw}$. We obtain the fidelity and latency as performance metrics. In all cases, it is assumed that the qubit of chip $A$ is to be transmitted, that it is initially in the excited state, and that the transmission starts at $t=0$. Then, the fidelity is calculated as the maximum probability of the qubit of chip $B$ to be in the excited state, and the latency is the point in time when that happens. 

The study is divided in two main cases, which we name as \textit{resonant} and \textit{out of resonance}. First, as baseline, we study the case where all elements resonate at the same frequency ($\omega_q = \omega_w = 6$ GHz) and hence focus on the effect of the decay rates and coupling strength on the response. Then, we study the impact of the detuning $\Delta = |\omega_q - \omega_w |$ by simulating a system where the waveguide resonates at a different frequency than the qubits. In particular, our analysis will consider waveguide frequencies greater than the qubit frequency. Higher-frequency waveguides are advantageous due to their compact dimensions, making them well-suited for on-chip integration and scalability.


\begin{table}[!b]
    \centering
    \caption{Waveguide-qubit system parameters.}\vspace{-0.1cm}
    \label{tab:parameters}
    \begin{tabular}{|c|c|c|}
        \hline
        Parameters & Resonant & Out of resonance \\ 
        \hline \hline
        $\omega_w$   & 6 GHz  & 7 GHz to 50 GHz   \\ 
        \hline 
        $\kappa$   & \multicolumn{2}{c|}{1 MHz to 1 GHz} \\
        \hline \hline
        $\omega_q$   & \multicolumn{2}{c|}{6 GHz}   \\ 
        \hline
        $\gamma$   & \multicolumn{2}{c|}{1 MHz to 1 GHz} \\
        \hline \hline 
        $g_{qw}$   &  \multicolumn{2}{c|}{50 MHz to 1 GHz} \\ 
        \hline   
    \end{tabular}
\end{table}

\section{Results}
\label{sec:results}
The fidelity and latency of the proposed qubit-waveguide system is evaluated in this section. We first analyze the resonant (or tuned) system in Sec.~\ref{sec:tuned} and then the out-of-resonance (or detuned) system in Sec.~\ref{sec:detuned}

\subsection{Resonant qubit-waveguide system} \label{sec:tuned}
Fig. \ref{lossless-config} illustrates the time evolution of the tuned qubit-waveguide system, assuming negligible losses and decoherence ($\kappa \approx \gamma \approx 0$). In this case, the quantum state exchange fidelity is 100\%, whereas the photon transmission duration to the target qubit in chip B depends on the coupling strength factor mentioned in Eq. \eqref{Ht}. To illustrate this, Fig. \ref{g-increased-config} showcases a faster photon exchange between the two qubits as we increase the qubit-waveguide coupling strength $g_{qw}$ to 0.1 GHz. Effectively, the latency is cut to half as the coupling strength is doubled, suggesting a direct inverse relationship between the two factors in this simplified scheme devoid of second-order effects.  

In realistic experimental qubit-waveguide systems, the qubit decay rate $\gamma$ and the waveguide decay rate $\kappa$ disrupt inter-core communications and operations. To illustrate this, we introduce the system losses through the Lindblad master equation, which affects the transmitted state fidelity. As we show in Fig. \ref{lossy-config}, even with a relatively low loss ($\kappa = \gamma = 1$ MHz), the amplitude of the state exchange oscillation is reduced over time. The continuously damped oscillations, referring to the photon exchange over time, indicates a loss of the quantum state due to cumulative dissipating effect of the noise factors after each exchange cycle. 

Next, we analyze the effect of the decay factors on the fidelity and latency of the photon transmission operation more broadly. Figs. \ref{resonant-waveguide-decay resonant} and \ref{resonant-qubit-decay resonant} show the degrading impact of system losses on the state fidelity of photon transfer operations. From the comparison of the two figures, we observe that the qubit decay rate is more important since the occupation probability is generally lower in the waveguide than in the qubit sites.

Another aspect worth noting from Figs. \ref{resonant-waveguide-decay resonant} and \ref{resonant-qubit-decay resonant} is that the qubit-waveguide coupling strength $g_{qw}$ is a crucial factor for the efficiency of the state transfer. In fact, strong coupling factors can compensate for the effect of system losses. This is directly related to the latency of the state exchange; as mentioned above, stronger coupling reduces the time required to achieve the quantum state transfer and therefore reduces the likelihood of the qubit decohering or the photon being lost in the waveguide. Fig. \ref{latency-resonant} plots the latency of the state transfer as a function of the coupling factor, hence highlighting the significant role of robust qubit-waveguide coupling to guarantee a reliable state transfer.

\begin{figure*}[t]
    \centering
    \begin{subfigure}[t]{0.34\textwidth}
        \centering
        \includegraphics[width=\columnwidth]{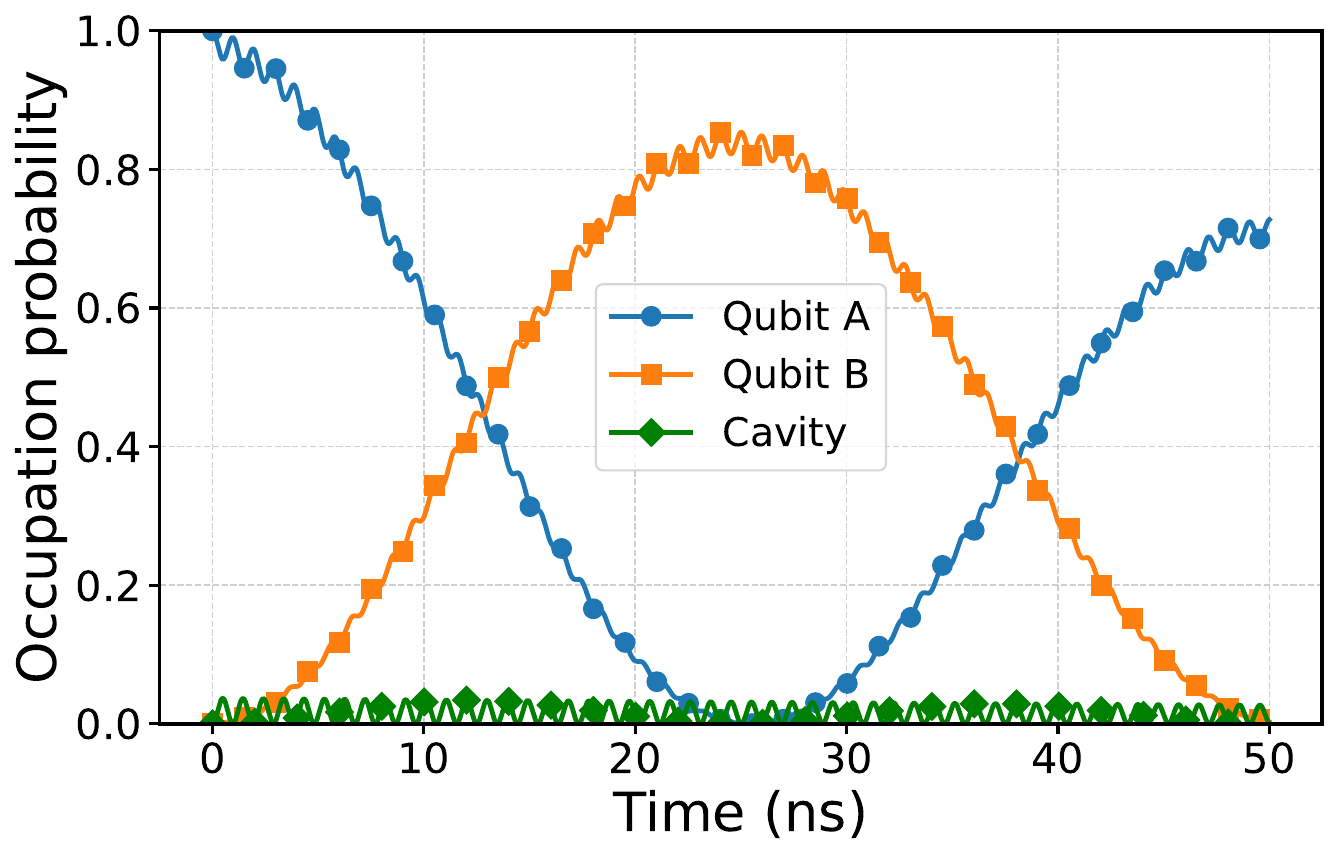} 
        \caption{\centering Lossy setup with $\kappa = \gamma = 1$ MHz, $g_{qw}= 0.1$~GHz, and $\omega_w = 7$ GHz.}
       \label{detuned 7GHz} 
    \end{subfigure}
    \hfill
    \begin{subfigure}[t]{0.34\textwidth}
        \centering
        \includegraphics[width=\columnwidth]{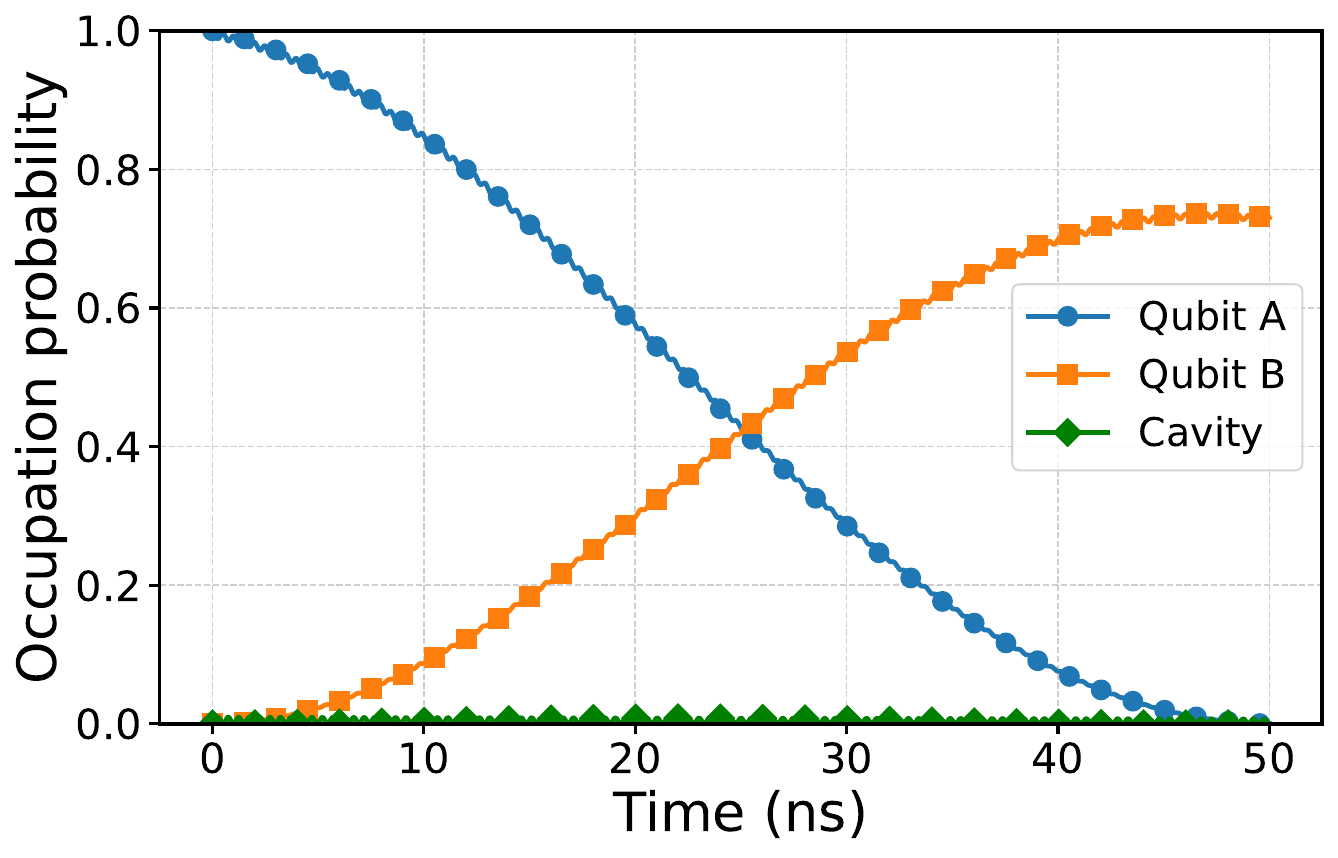} 
        \caption{\centering Lossy setup with $\kappa = \gamma = 1$ MHz, $g_{qw}= 0.1$~GHz, and $\omega_w = 8$ GHz.}
    \label{Detuned 8 GHz}    
    \end{subfigure}
     \hfill
    \begin{subfigure}[t]{0.3\textwidth}
         \centering
    \includegraphics[width=\columnwidth]{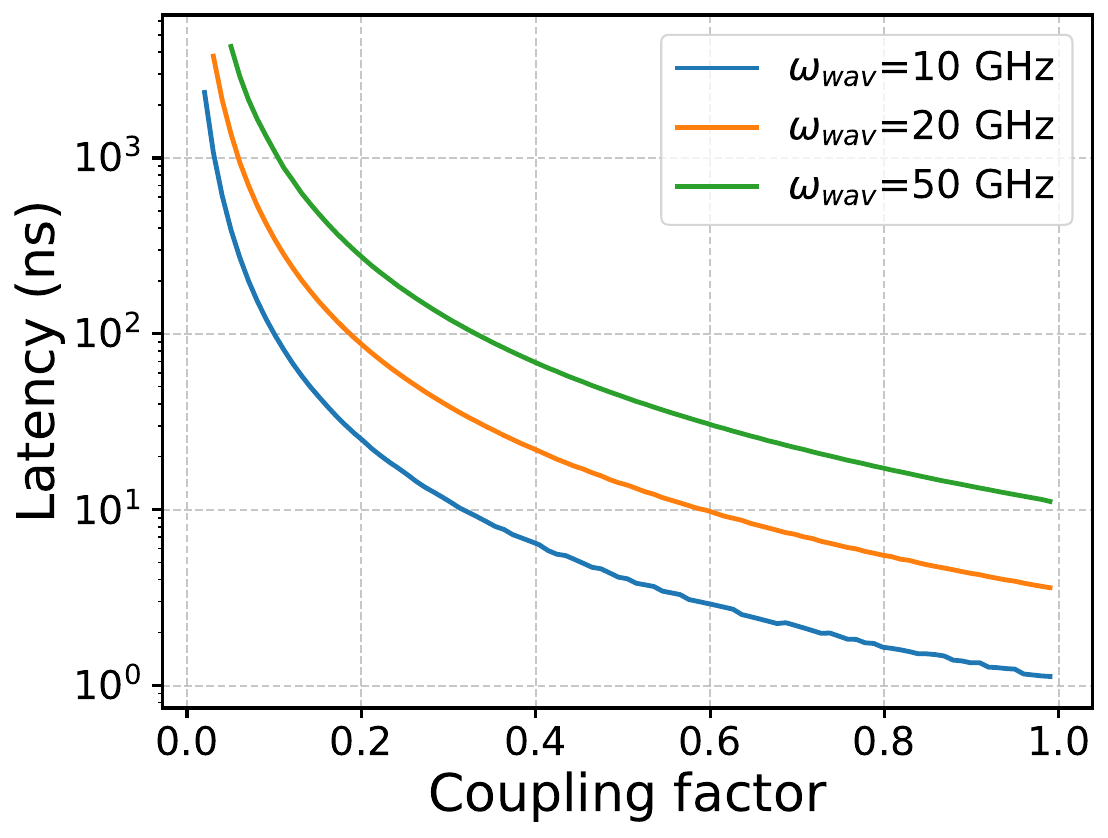}
    \caption{\centering Latency as a function of the coupling strength for three detuning values.}
    \label{latency and length}
    \end{subfigure}   
    \caption{Time-domain analysis of the quantum state transfer between two qubits in the out-of-resonance case ($\omega_q \neq \omega_w$).}
    \label{fig:Time evolution and state manipulation (non-resonant prototype)}
\end{figure*}

\begin{figure*}[t]
    \centering
    \begin{subfigure}[t]{0.32\textwidth}
        \centering
        \includegraphics[width=\columnwidth]{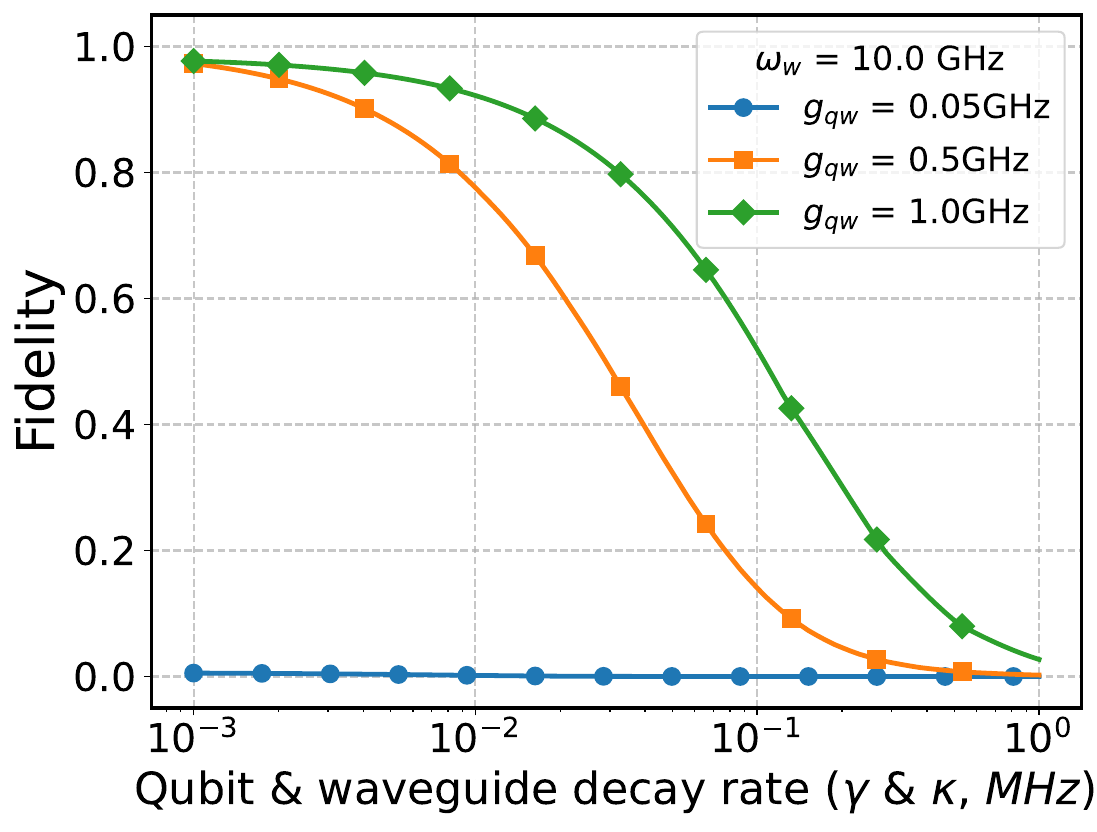}
        \caption{\centering Fidelity over losses and coupling factor for $\omega_w = 10$ GHz.}
       \label{losses 10GHz} 
    \end{subfigure}
    \hfill
    \begin{subfigure}[t]{0.32\textwidth}
        \centering
        \includegraphics[width=\columnwidth]{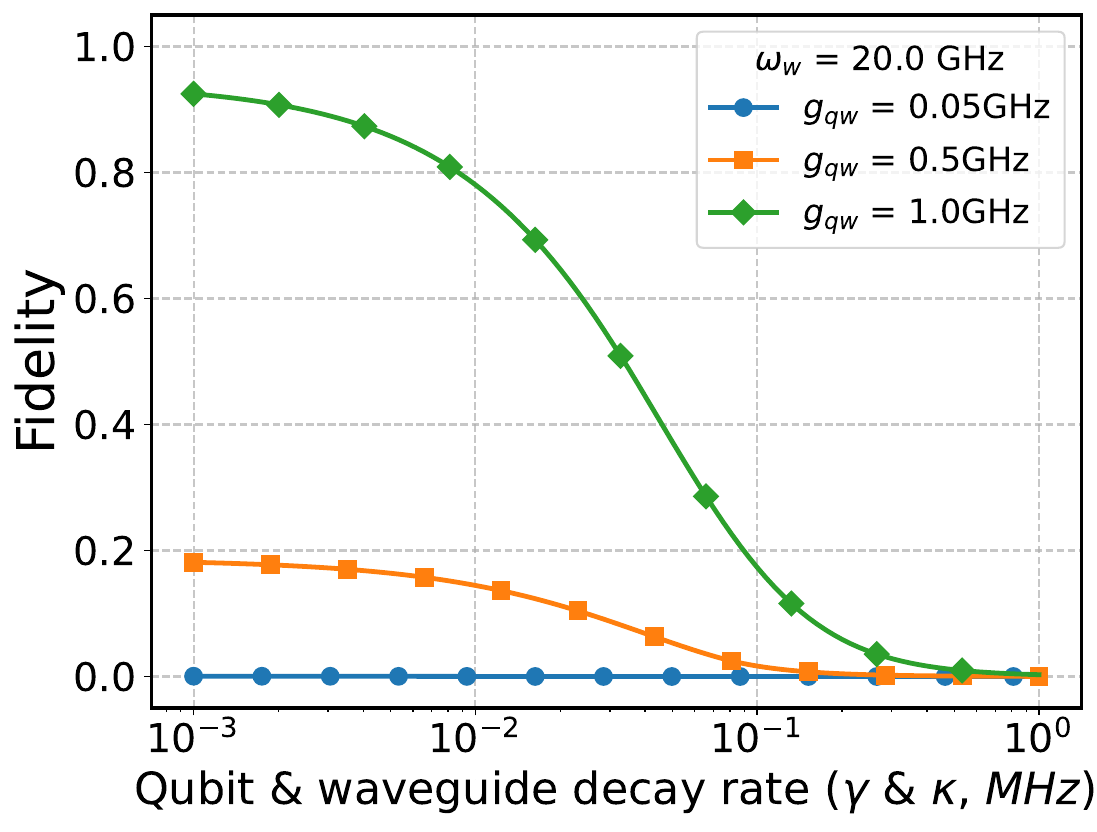}
        \caption{\centering Fidelity over losses and coupling factor for $\omega_w = 20$ GHz.}
    \label{losses 20GHz}    
    \end{subfigure}
     \hfill
    \begin{subfigure}[t]{0.32\textwidth}
        \centering
        \includegraphics[width=\columnwidth]{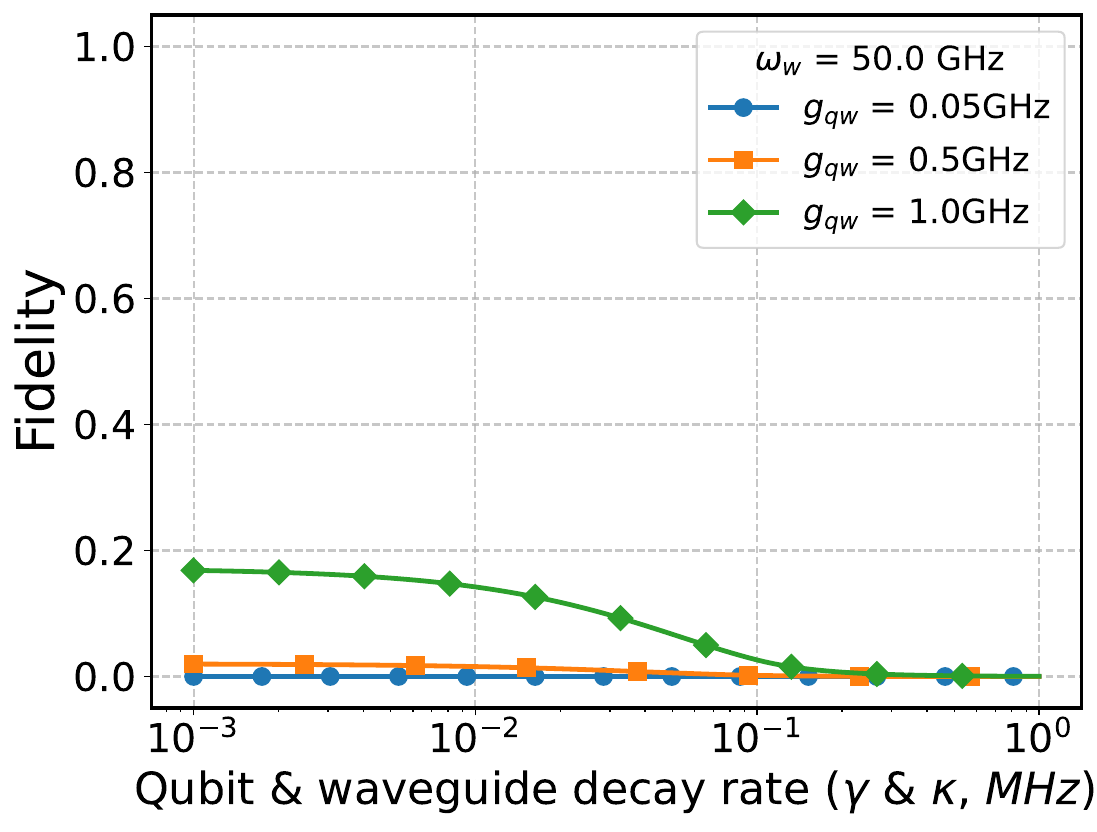}
        \caption{\centering Fidelity over losses and coupling factor for $\omega_w = 50$ GHz.}
    \label{losses 50GHz}    
    \end{subfigure}   
    \caption{Assessment of the quantum state transfer performance in the out-of-resonance case ($\omega_q \neq \omega_w$).}\vspace{-0.2cm}
    \label{system losses non-resonant}
\end{figure*}

\subsection{Out-of-resonance qubit-waveguide system} \label{sec:detuned}
Given a lossy system, characterized by the decay factors $\kappa$ and $\gamma$ and the qubit decoherence, we estimate the quantum state transfer fidelity between qubits via a waveguide under detuned conditions. In Figs. \ref{detuned 7GHz} and \ref{Detuned 8 GHz}, compared to the resonant case of Fig. \ref{lossy-config}, we clearly observe a much slower coupling. In fact, the latency is proportional to the detuning as further observed in Fig. \ref{fig:Time evolution and state manipulation (non-resonant prototype)} because the effective coupling scales with $g_{eff} \sim g/\sqrt{\Delta^2 + \gamma^2}$ \cite{schuster2010high}, to the point of becoming one or even two orders of magnitude higher than the resonant case.

To further study the detuned system, we vary the waveguide frequency $\omega_w$ from 10 to 50 GHz and the coupling strength $g_{qw}$ from 0.05 to 1 GHz, while sweeping the qubit and waveguide losses. The results, summarized in Fig.~\ref{system losses non-resonant}, show how the fidelity is affected by losses in a non-linear way. Moreover, the fidelity is evidently impacted by the detuning factor due to the sharp increase in the latency of the entire operation. Hence, in such a detuned system, a high coupling factor is crucial to compensate for the impact of detuning.

\section{Discussion}
\label{sec:discussion}
In this work, we examine tuned and detuned qubit-waveguide systems, focusing on their practicality for interconnecting quantum processors. Tuned frequency systems are ideal for reliable waveguide-based interconnects, offering high state transmission fidelity and low latency, especially with strong coupling. Strong coupling suppresses the impact of system losses, such as waveguide decay rate $\kappa$ and qubit decay rate $\gamma$. However, the larger physical size of resonant waveguides poses challenges for integrating nanometric qubits and small processors in cryogenic environments.

We also explore detuned systems to understand the effects of waveguide frequency detuning. In this case, the results suggest that a qubit with long decoherence time is more important than a low-loss waveguide, due to the low effective coupling factor (and hence low probability of having a photon enclosed in the waveguide). This seems to imply that waveguides with high coupling factor may be preferred, even if that implies a higher intrinsic loss -- because strong qubit-waveguide coupling enhances robustness against losses and decoherence. 

In this context, a third factor is the maximum detuning factor that can be admissible without disrupting the quantum state transfer. Higher frequencies lead to more compact waveguides (whose dimensions scale with the wavelength) yet at the expense of lowering the effective coupling rate, and therefore the overall fidelity. Moreover, this complicates the design of the qubit-waveguide couplers. 

In summary, there is a fundamental and interesting tradeoff between losses, coupling, and detuning worth exploring in future work. This tradeoff is well encapsulated in the expression of the quality factor $Q$ of the entire system \cite{schuster2010high}, which is 
\begin{equation}
    Q = \frac{\Delta^2 + \gamma^2}{2g_{qw}^2\gamma + \kappa(\Delta^2 + \gamma^2)} \omega_w
\end{equation}
and therefore illustrates a non-trivial search for optimal design parameters that maximizes fidelity and minimizes latency. 


\section{Conclusion}
\label{sec:conclusion}
This work highlights the promising role of waveguide-based channels in modular quantum computing. Using the waveguide QED formalism, we model both resonance and out-of-resonance configurations. Tuned systems with strong coupling remain the prime choice to maximize fidelity. However, the large size of low-frequency waveguides presents severe integration challenges. For highly detuned and miniaturized waveguides, achieving high fidelity demands strong coupling strength, which is fundamentally difficult in these systems.

\bibliographystyle{IEEEtran}
\bibliography{references}
\end{document}